\magnification1095
\input amstex
\hsize32truecc
\vsize44truecc
\input amssym.def
\font\ninerm=plr9 at9truept
\font\twbf=cmbx12
\font\twrm=cmr12

\footline={\hss\ninerm\folio\hss}

\def\rf#1 {\text{(#1)}}
\def\section#1{\goodbreak \vskip20pt plus5pt
\noindent {\bf #1}\vglue4pt}
\def\eqn#1 {\eqno(\text{\rm#1})}
\let\al\aligned
\let\eal\endaligned
\let\ealn\eqalignno
\def\cL{\Cal L}
\let\o\overline
\let\ul\underline
\let\a\alpha
\let\b\beta
\let\d\delta
\let\e\varepsilon
\let\D\Delta
\let\g\gamma
\let\G\varGamma
\let\la\lambda

\let\z\zeta
\let\pa\partial
\let\t\widetilde
\let\u\tilde
\def\uP{\ul{\t P}{}}
\def\RG{\t R(\t\G)}
\def\arctg{\operatorname{arctg}}
\def\ctg{\operatorname{ctg}}
\def\tg{\operatorname{tg}}
\def\Li{\operatorname{Li}}
\def\Re{\operatorname{Re}}
\def\br#1#2{\overset\text{\rm #1}\to{#2}}
\def\({\left(}\def\){\right)}
\def\[{\left[}\def\]{\right]}
\def\dr#1{_{\text{\rm#1}}}
\def\gi#1{^{\text{\rm#1}}}
\def\ct{constant}
\def\co{cosmological}
\def\dt{dependent}
\def\ef{effective}
\def\ex{exponential}
\def\lg{lagrangian}
\def\gr{gravitational}
\def\wrt{with respect to }
\def\dint{-\kern-11pt\intop}
\def\rAB{\text{AB}}
\def\Int{\operatorname{int}\nolimits}
\def\crt{\operatorname{crt}\nolimits}
\def\q{quintessence}
\let\P\varPsi

{\twbf
\advance\baselineskip4pt
\centerline{M. W. Kalinowski}
\centerline{\twrm (Higher Vocational State School in Che\l m, Poland)}
\centerline{A Warp Factor}
\centerline{in the Nonsymmetric}
\centerline{Kaluza--Klein (Jordan--Thiry) Theory}}

\vskip20pt plus5pt

{\bf Abstract.} We consider in the paper some consequences of the
Nonsymmetric Kaluza--Klein (Jordan--Thiry) Theory with spontaneous symmetry
breaking connecting to the existence of the warp factor.

\vskip20pt plus5pt
We develop in the paper some applications and consequences of the
Nonsymmetric Kaluza--Klein (Jordan--Thiry) Theory extensively presented in
the first point of Ref.~[1]. We refer for all the details to Ref.~[1],
especially to the book on nonsymmetric fields theory and its applications. 

Let us give a short description of the theory.

We  develop  a  unification  of  the  Nonsymmetric
Gravitational Theory and gauge fields  (Yang--Mills'  fields)  including
spontaneous symmetry breaking and  the  Higgs'  mechanism  with  scalar
forces  connected  to  the  gravitational  constant.   The  theory  is 
geometric and unifies tensor-scalar gravity with massive  gauge  theory 
using a multidimensional manifold in a Jordan--Thiry manner.
We use a nonsymmetric version of this theory.
The general scheme is the following.  We  introduce
the principal fibre bundle over  the  base $V = E \times  G/G_{0}$  with  the
structural group $H$, where $E$ is a space-time, $G$ is a compact  semisimple
Lie group, $G_{0}$ is its compact subgroup and $H$  is  a  semisimple  compact
group.  The manifold $M = G/G_{0}$ has an interpretation as a ``vacuum states
manifold" if $G$ is broken to $G_{0}$ (classical vacuum states).  We define on
the space-time $E$, the nonsymmetric tensor $g_{\alpha\beta}$  from  N.G.T.,  which  is
equivalent to the existence of two geometrical objects
$$\eqalignno{\noalign{\vskip2pt}
\overline{g}&= g_{(\alpha \beta )}\overline{\theta }^\alpha \otimes \overline{\theta}^\beta \cr
\noalign{\vskip2pt}
\underline{g}&= g_{[\alpha \beta ]}\overline{\theta }^\alpha \wedge \overline{\theta}^\beta
\cr\noalign{\vskip2pt}}
$$
the symmetric tensor $\overline{g}$ and the  2-form $\ul{g}$.  Simultaneously we  introduce
on $E$ two connections from N.G.T. $\overline{W}_{\beta \gamma}^\alpha $ and 
$\widetilde{\overline{\varGamma}}_{\beta \gamma}^\alpha $. On the homogeneous space $M$
we define the nonsymmetric metric tensor 
$$
g_{\tilde{a}\tilde{b}} = h^{0}_{\tilde{a}\tilde{b}} 
+ \zeta  k^{0}_{\tilde{a}\tilde{b}} 
$$
where $\zeta $ is the dimensionless constant, in  a  geometric  way.  Thus  we
really have the nonsymmetric metric tensor on or $V = E \times 
G/G_{0}$. 
$$
\gamma _{\rAB} =
\left(\vcenter{\offinterlineskip\tabskip0pt
\halign{\strut\tabskip5pt plus10pt minus10pt
\hfill#\hfill&\vrule height10pt#&\hfill#\hfill\tabskip0pt\cr
$\vrule height0pt depth6pt width0pt g_{\alpha \beta }$ && $0$\cr 
\noalign{\hrule}
$\vrule height9pt depth4pt width0pt 0$ &&
$r^2g_{\tilde{a}\tilde{b}}$\cr}}\right) 
$$
$r$ is a parameter which characterizes the size   of  the  manifold $M =
G/G_{0}$.  Now on the principal bundle $\ul{P}$ we define the connection $\omega $,  which
is the 1-form with values in the Lie algebra of $H$.

After  this  we   introduce   the   nonsymmetric   metric   on $\ul{P}$
right-invariant with respect to the action of the group $H$,  introducing
scalar field $\rho $ in a Jordan--Thiry manner.
The only  difference  is  that
now  our  base  space  has   more   dimensions   than   four.   It   is
$(n_{1}+4)$-dimensional, where $n_{1} = \dim (M) = \dim (G)-\dim (G_{0})$.  In  other
words,  we  combine  the  nonsymmetric  tensor $\gamma _{\rAB}$  on $V$   with   the
right-invariant nonsymmetric tensor on the group $H$ using the connection
$\omega $ and the scalar field $\rho $.  We suppose that the factor $\rho $  depends  on  a
space-time point only.  This condition can be abandoned and we consider
a more general case where $\rho =\rho (x,y)$, $x\in E$, $y\in M$ resulting in  a  tower  of
massive scalar field $\rho _{k}, k=1,2\ldots .$  This  is  really  the Jordan--Thiry
theory  in  the  nonsymmetric  version  but   with $(n_{1}+4)$-dimensional
``space-time".  After this we act in the classical manner.
We introduce the linear  connection
which is compatible with this nonsymmetric metric. This  connection  is
the multidimensional analogue of the connection
$\widetilde{\overline{\varGamma}}{}^{\alpha}_{\beta \gamma }$ on the  space-time
$E$.  Simultaneously  we  introduce  the  second  connection   $W$.    The
connection $W$ is the multidimensional analogue of the $\overline{W}$-connection  from
N.G.T. and Einstein's Unified Field Theory.  
Now we  calculate  the  Moffat--Ricci
curvature scalar $R(W)$ for the connection $W$ and  we  get  the  following
result. $R(W)$ is equal to the sum of the Moffat--Ricci curvature on  the
space-time $E$ (the  gravitational  lagrangian  in  Moffat's  theory  of
gravitation), plus $(n_{1}+4)$-dimensional lagrangian  for  the  Yang--Mills'
field from the Nonsymmetric Kaluza--Klein Theory plus  the  Moffat--Ricci
curvature scalar on the homogeneous  space $G/G_{0}$  and  the Moffat--Ricci
curvature scalar on the group $H$ plus  the  lagrangian  for  the  scalar
field~$\rho $.  The only difference is that our Yang--Mills' field is  defined
on $(n_{1}+4)$-dimensional  ``space-time"   and   the   existence   of   the
Moffat--Ricci
curvature scalar of the connection on the homogeneous space
$G/G_{0}$.
All of these terms (including $R(\overline{W})$)  are  multiplied  by  some
factors depending on the scalar field~$\rho $.

This lagrangian depends on the point of $V = E \times  G/G_{0}$ i.e.\ on  the
point of the space-time $E$ and on the point 
of $M = G/G_{0}$.  The  curvature
scalar on $G/G_{0}$ also depends on the point of $M$.

We now go to the group structure  of  our  theory.   We  assume $G$
invariance of the connection $\omega $ on the principal fibre bundle $\ul{P}$, the so
called Wang-condition.   According  to  the
Wang-theorem the  connection $\omega $  decomposes  into  the
connection $\widetilde{\omega }\dr{E}$  on  the  principal  bundle $Q$  over  space-time $E$  with
structural group $G$ and the multiplet of scalar fields ${\varPhi} $.  Due  to  this
decomposition the multidimensional  Yang--Mills'  lagrangian  decomposes
into: a 4-dimensional Yang--Mills' lagrangian with  the  gauge  group $G$
from the Nonsymmetric Kaluza--Klein Theory, plus  a  polynomial  of  4th
order with respect to the fields ${\varPhi} $, plus a term which is quadratic with
respect to the gauge derivative of ${\varPhi}$  (the gauge derivative with respect
to the connection $\widetilde{\omega }\dr{E}$ on a space-time $E$) plus a new term which is of 2nd
order in the ${\varPhi} $, and is linear with respect  to  the Yang--Mills'  field
strength.  After this we perform the  dimensional  reduction  procedure
for the Moffat--Ricci scalar curvature on the manifold $\ul{P}$.   We  average
$R(W)$ with respect to the homogeneous space $M = G/G_{0}$.
In this way we get the lagrangian of our theory.  It is the
sum  of  the  Moffat--Ricci  curvature  scalar   on $E$ (gravitational
lagrangian) plus a Yang--Mills' lagrangian with gauge group $G$  from  the
Nonsymmetric Kaluza--Klein Theory plus a  kinetic  term  for
the scalar field ${\varPhi} $, plus a potential $V({\varPhi} )$ which is of  4th  order  with
respect to ${\varPhi} $,  plus ${\Cal L}_{\Int}$  which  describes  a  nonminimal  interaction
between the scalar field ${\varPhi} $ and the Yang--Mills' field, plus cosmological
terms, plus  lagrangian  for  scalar  field $\rho $.   All  of  these  terms
(including $\overline{R}(\overline{W})$) are multiplied of course by some factors depending on
the scalar field $\rho $.  We redefine tensor $g_{\mu \nu }$ and $\rho$
and pass from scalar field $\rho $ to ${\varPsi} $
$$ 
\rho  = e^{-{\varPsi} }
$$
After this  we  get  lagrangian  which  is  the  sum  of  gravitational
lagrangian,   Yang--Mills'   lagrangian,   Higgs'   field    lagrangian,
interaction  term ${\Cal L}_{\Int}$  and  lagrangian  for  scalar  field ${\varPsi} $   plus
cosmological terms.  These terms depend now on the scalar field ${\varPsi} $.   In
this way we have in our theory a multiplet of scalar fields $({\varPsi} ,{\varPhi} )$.   As
in  the  Nonsymmetric-Nonabelian   Kaluza--Klein   Theory   we   get   a
polarization  tensor  of  the  Yang--Mills'   field   induced   by   the
skewsymmetric part of the metric on the space-time and on the group $G$.
We get an additional term in the Yang--Mills' lagrangian induced by  the
skewsymmetric part of the metric $g_{\alpha \beta }$.  We get also ${\Cal L}_{\Int}$,
which is absent in the dimensional reduction procedure known up to  now.
Simultaneously,  our  potential  for  the
scalar--Higgs' field really differs from the analogous potential.
Due to the skewsymmetric part of the metric  on $G/G_{0}$
and on $H$ it has a more complicated structure.   This  structure  offers
two kinds of critical points for the minimum of  this  potential: ${\varPhi} ^{0}_{\crt }$
and ${\varPhi} ^{1}_{\crt }$.  The first is known in the classical,  symmetric  dimensional
reduction procedure and  corresponds  to  the  trivial
Higgs' field (``pure gauge").  This is the ``true" vacuum  state  of  the
theory.  
The second, ${\varPhi} ^{1}_{\crt }$, corresponds to a more complex configuration.
This is only a local (no absolute) minimum  of~$V$. It is  a  ``false"
vacuum. The Higgs' field is not a ``pure" gauge here.  In the first case
the unbroken group is always $G_{0}$.  In the second case, it is in  general
different and strongly depends on the details of the theory: groups $G_{0}$, 
$G$, $H$, tensors $\ell _{ab}$, $g_{\tilde{a}\tilde{b}}$ and  the  constants $\zeta$, $\xi $.   It  results  in  a
different spectrum of mass for intermediate bosons.  However, the scale
of the mass is the same and it is fixed by a constant $r$ (``radius"  of
the manifold $M = G/G_{0})$.  In the first case $V({\varPhi} ^{0}_{\crt }) =0$, in  the  second
case it is, in general, not zero $V({\varPhi} ^{1}_{\crt }) \neq 0$.  Thus, in the first case,
the cosmological constant is a sum of the scalar  curvature  on $H$  and
$G/G_{0}$, and in the second case, we should  add  the  value $V({\varPhi} ^{1}_{\crt })$.   We
proved that using the constant $\xi$ we are able in some cases to make  the
cosmological constant as small as we want (it  is  almost  zero,  maybe
exactly zero, from the observational data point of view).  Here we  can
perform the same procedure for the  second  term  in  the  cosmological
constant using the constant $\zeta $.  In the first case we are able  to  make
the cosmological constant sufficiently small but this is  not  possible
in general for the second case.

The transition from ``false" to ``true" vacuum occurs  as  a  second
order phase transition.  We discuss this transition  in
context of the first order phase transition in models of  the  Universe.
In this paper the interesting  point  is  that  there
exists an effective scale of masses, which depends on the scalar 
field~${\varPsi} $.

Using Palatini variational principle we get an equation for fields
in our theory.  We find  a  gravitational  equation  from  N.G.T.  with
Yang--Mills', Higgs' and  scalar  sources  (for  scalar  field ${\varPsi} )$  with
cosmological terms.  This gives us  an  interpretation  of  the  scalar
field ${\varPsi} $ as an effective gravitational constant
$$
G\dr{eff} = G_{N}e^{-(n+2){\varPsi} }
$$
We get an equation for this scalar  field ${\varPsi} $.   Simultaneously  we  get
equations for Yang--Mills' and Higgs' field. We also discuss the  change
of the effective scale of mass, $m\dr{eff}$ with a relation to the  change  of
the gravitational constant~$G\dr{eff}$.

In the ``true" vacuum case we  get  that  the  scalar  field ${\varPsi} $  is
massive  and  has  Yukawa-type  behaviour.   In  this  way   the   weak
equivalence principle is satisfied.  In the  ``false"  vacuum  case  the
situation is more complex.  It  seems  that  there  are  possible  some
scalar forces with infinite range.  Thus  the  two  worlds  constructed
over the ``true" vacuum and the ``false" vacuum  seem  to  be  completely
different: with different unbroken groups, different mass spectrum  for
the broken gauge and Higgs' bosons,  different  cosmological  constants
and with different behaviour for the scalar field ${\varPsi} $.   The  last  point
means that in the ``false" vacuum case the  weak  equivalence  principle
could be violated and the gravitational  constant  (Newton's  constant)
would increase in distance between bodies.

We explore Einstein $\la$-transformation for a connection $W^{\u A}_{\u B}$
($(m+4)$-dimensional) in order to get an interpretation of ${\Bbb R}_+$ gauge
invariance for a field $\t W_\mu$.
We discuss ${\Bbb R}_{+}$ and ${\bold U}(1)\dr{F}$ invariance.
We  decide  that
${\bold U}(1)\dr{F}$  invariance  from  G.U.T.  is  a  local  invariance.   Due  to  a
geometrical construction we are  able  to  identify  $\overline{W}_{\mu }$  from  Moffat's
theory of gravitation with the four-potential $\widetilde{A}\gi{F}_{\mu }$ corresponding  to  the
${\bold U}(1)\dr{F}$ group (internal rotations connected to fermion charge).  In  this
way, the fermion number is conserved and plays the role of  the  second
gravitational charge.  Due to the Higgs' mechanism
$\widetilde{A}\gi{F}_{\mu }$  is  massive  and
its strength, $\widetilde{H}\gi{F}_{\mu \nu }$, is of short range with Yukawa-type behaviour.   This
has  important  consequences.   The   Lorentz-like   force   term   (or
Coriolis-like force term) in the equation of motion for a test particle
is of short range with Yukawa-type behaviour.  The range of this  force
is smaller than the  range  of  the  weak  interactions.   Thus  it  is
negligible in the equation of motion for a test particle.   We  discuss
the possibility of the cosmological origin of the mass  of  the  scalar
field ${\varPsi} $ and geodetic equations on $\ul{P}$.  We consider an infinite tower  of
scalar fields ${\varPsi} _{k}(x)$ coming from the expansion of the  field ${\varPsi} (x,y)$  on
the manifold $M=G/G_{0}$ 
into harmonics of  the  Beltrami--Laplace  operator.
Due to Friedrichs' theory we can 
diagonalize  an  infinite  matrix  of
masses for ${\varPsi} _{k}$ transforming them into new  fields
${\varPsi}'_k$.   The  truncation 
procedure means here to take a zero mass mode 
${\varPsi} _{0}$ and equal it to ${\varPsi} $.

We consider a possibility to take seriously additional dimensions in our
theory in a framework similar to a Randall--Sundrum scenario. In our theory
the additional dimensions connecting to the manifold~$M$ (a~vacuum manifold)
could be considered in such a way. They are not directly observable because
the size of the manifold is very small. In order to see them it is necessary
to excite massive modes of the scalar field (a~tower of those fields). We
find an interesting toy model (a~5-dimensional model) which can describe a
possibility to travel with speed higher than a speed of light using the fifth
dimension. This dimension has nothing to do with the fifth dimension in
Kaluza--Klein theory. We do some analysis on an energy to excite a scalar
field~$\varPsi$ to get this special solution to the theory. We discuss also
some quantitative relations involving travelling signals in the model via
fifth dimension. We consider various posibilities to excite a warp factor due
to fluctuations of a tower of scalar fields finding a density of an
excitation energy. Eventually we find a zero energy (or almost zero)
condition for such an excitation. We consider also a simple solution for
hierarchy problem in the framework of the Nonsymmetric Kaluza--Klein
(Jordan--Thiry) Theory. 

Let us consider the Eqs (5.3.17--19) of the first point of Ref.~[1], p.~329, 
and let us release the condition that
$\rho$ is in\dt\ of~$y$. Thus $\rho=\rho(x,y)$, $y\in G/G_0$. In
Eq.~(5.3.34) of the first point of Ref.~[1], p.~333, we get in a place of
$$
\frac{\la^2}4 \rho^{n-2} \(\o M \t g^{(\g\mu)}\rho_{,\g}\rho_{,\mu}
+n^2g^{[\mu\nu]}g_{\d\mu}\t g^{(\d\g)}\)\rho_{,\nu}\cdot \rho_{,\g}
\eqn1
$$
the formula
$$
\frac{\la^2}4 \rho^{n-2} \(\o M \t \g^{(CM)}\rho_{,C}\rho_{,M}
+n^2\g^{[MN]}\g_{DM}\t \g^{(DC)}\rho_{,N}\cdot \rho_{,C}\).
\eqn2
$$
This $\rho$ has nothing to do with a density of energy considered below.

Now let us repeat the procedure from Section~5.5 of the first point of
Ref.~[1], i.e. the redefinition of $g_{\mu\nu}$ and~$\rho$. We get the
formula (5.5.5) of the first point of Ref.~[1], p.~355, but in a place of
$\cL\dr{scal}(\varPsi)$ we get the formula (5.14.3). Simultaneously
$\varPsi=\varPsi(x,y)$, $x\in E$, $y\in G/G_0$ and the metric on a space-time
$E$ depends on $y\in G/G_0$, i.e.
$$
g_{\mu\nu}=g_{\mu\nu}(x,y), \quad x\in E, \quad y\in G/G_0. \eqn3
$$
Thus $g_{\mu\nu}$ is parametrized by a point of $G/G_0$. Simultaneously we
can interpret a dependence on higher dimensions as an existence of a tower of
scalar fields $\varPsi_K$ (see Eqs (5.14.4--8), Eqs (5.14.11--14) from
the first point of Ref.~[1], p.~434--436, and a discussion
below).

The interesting point will be to find physical consequences of this
dependence for~$g_{\mu\nu}$. This can be achieved by considering \co\
solutions of the theory. Thus let us come to Eq.\ (5.5.5) of the first point
of Ref.~[1], p.~355, supposing the \lg\ of
matter fields is written as $L\dr{matter}$, $g_{\mu\nu}=g_{\nu\mu}$ and
depends on $y\in G/G_0$. We get
$$
\al
L\sqrt{-g}\sqrt{|\t g|} &= \sqrt{-g}\sqrt{|\t g|}
\biggl(\o R(\t{\o \G})+\frac{\la^2}4 e^{-(n+2)\varPsi} L\dr{matter}\\
&+\frac{\la^2}4 \cL\dr{scal}(\varPsi)+ \frac1{\la^2}e^{(n+2)\varPsi}\t R(\t \G)
+\frac1{r^2}e^{n\varPsi}\ul{\t P}\biggr).
\eal \eqn4
$$

According to the standard interpretation of the constant $\la$ we have
$$
\frac{\la^2}4 \sim \frac1{m^2\dr{pl}} \eqn5
$$
and
$$
\frac1{r^2}\sim m^2_{\u A} \eqn6
$$
where $m_{\u A}$ is a scale of a mass of broken gauge bosons in our theory
and $m\dr{pl}$ is a Planck mass, $c=1$, $\hbar=1$.

Thus one gets form variation principle with respect to $g_{\mu\nu}$
and~$\varPsi$ 
$$
\al
\t{\o R}_{\mu\nu} - \frac12 \t{\o R}g_{\mu\nu}&=
\frac{e^{-(n+2)\varPsi}}{m^2\dr{pl}}\,
\br{matter}T_{\kern-9pt \mu \nu }+\frac1{m^2\dr{pl}}
\br{scal}T_{\kern-5pt\mu \nu }\\
&+\(\frac{m^2\dr{pl}}8 e^{(n+2)\varPsi}\t R(\t \G)+m^2_Ae^{n\varPsi}\ul{\t P}\)g_{\mu\nu}
\eal
\eqn7
$$
$$
\al
\o Mg^{\mu\nu}\t{\o \nabla}_\mu \t{\o\nabla}_\nu\varPsi
&+\frac1{r^2}\t L\varPsi + \frac{m\dr{pl}^4}4 (n+2)e^{(n+2)\varPsi}\t R(\t\G)\\
&+(m_{\u A}m\dr{pl})^2n e^{n\varPsi}\ul{\t P}-(n+2)Te^{-(n+2)\varPsi}=0
\eal \eqn8
$$
where $\t{\o R}_{\mu\nu}$ and $\t{\o R}$ are a Ricci tensor and a scalar
curvature of a Riemannian geometry induced by a metric $g_{\mu\nu}$ (which
can depend on $y\in G/G_0$).

$\t L$ is an operator on $G/G_0$ defined by Eq.\ (5.14.39) of the first point
of Ref.~[1], p.~442, $\t{\o\nabla}_\mu$
a covariant derivative with respect to a connection induced by $g_{\mu\nu}$
(a~Riemannian one).
$$
\br{matter}T^{\mu\nu}=(p+\varrho)u^\mu u^\nu- pg^{\mu\nu}. \eqn9
$$
Let us consider a simple model with a metric
$$
ds^2_4=e^{2v(y)}\,dt^2- d{\vec r}^{\,2} \eqn10
$$
and let us suppose that $G/G_0=S^2=SO(3)/SO(2)$. Thus
$$
ds^2_2=r^2\(d\chi^2+\sin^2\chi\,d\chi^2\), \eqn11
$$
$$
(\chi,\la)=y,\quad x\in \Bigl\langle0,\frac\pi2\Bigr),\quad \la\in\langle0,2\pi).
$$
In this case $v=v(\chi,\la)$ and
$$
g_{[\u a,\u b]}=g_{[56]}=-\zeta\sin\chi, \eqn12
$$
$\t L$ is defined by Eq.\ (5.14.41) of the first point of Ref.~[1], p.~443,
$$
\varPsi=\varPsi(\vec r,t,\chi,\la), \quad \varrho=\varrho(\vec r,t,\chi,\la),
\quad p=p(\vec r,t,\chi,\la).\eqn13
$$

Finally let us come to the toy model for which we suppose $\chi=\frac\pi2$
and we get \ef ly a 5-dimensional world $E\times S^1$, i.e.
$$
ds^2_5 = e^{2v(\la)}\,dt^2 - d{\vec r}^{\,2} - r^2\,d\la^2. \eqn14
$$
One easily gets
$$
\ealn{
R_{44}&=\(v''+(v')^2\)e^v &(15)\cr
R_{55}&=r^2\(v''+(v')^2\)e^{-v} &(16)
}
$$
where $'$ means a derivative with respect to $\la$. For simplicity we suppose
$\varrho=p=0$ and $\varPsi=\varPsi(\la)$ (does not depend on $\vec r$ and~$t$). In
this way
$$
\t L=\(\o M+\frac{n^2\zeta^2}{\zeta^2+1}\)\frac{\pa^2}{\pa\la^2}\,. \eqn17
$$
From Eq.\ (4) one obtains
$$
\(\frac{d^2}{d\la^2}e^v\)e^{-v} = \(\frac{m^2\dr{pl}}8 e^{(n+2)\varPsi}
\t R(\t \G)+ m^2_{\u A} e^{n\varPsi}\ul{\t P}\). \eqn18
$$
$$
\al
m^2_{\u A}\(\o M+\frac{n^2\zeta^2}{\zeta^2+1}\)\frac{d^2}{d\la^2}\varPsi
&+\frac{m\dr{pl}^4}4 (n+2)e^{(n+2)\varPsi}\t R(\t \G)\\
&+(m_{\t A}m\dr{pl})^2
ne^{n\varPsi}\ul{\t P}=0. 
\eal \eqn19
$$
The last equation can be transformed into
$$
\frac{d^2\varPsi}{d\la^2}=Ae^{(n+2)\varPsi}+Be^{n\varPsi} \eqn20
$$
where
$$
\ealn{
A&=-\frac{m^4\dr{pl}(n+2)}{4m^2_A(\o M+\frac{n^2\zeta^2}{\zeta^2+1})}
\t R(\t\G) &(21)\cr
B&=-\frac{m^2\dr{pl}n}{(\o M+\frac{n^2\zeta^2}{\zeta^2+1})}\ul{\t P}.
&(22)
}
$$
Supposing $\varPsi=\varPsi_0=\text{const.}$ one gets
$$
\ealn{
&Ae^{2\varPsi_0}+B=0 &(23)\cr
&e^{\varPsi_0}=\sqrt{-\frac BA}=\frac{4m_{\u A}}{m\dr{pl}}
\sqrt{\frac{n|\ul{\t P}|}{(n+2)\t R(\t\G)}}\,. &(24)
}
$$
Thus from Eq.\ (18) one gets ($z=e^v$)
$$
\frac{d^2z}{d\la^2}=\t Az \eqn25
$$
where
$$
\al
\t A&=\frac{m^2\dr{pl}}8 e^{(n+2)\varPsi_0}\t R(\t\G)+
m^2_A e^{n\varPsi_0}\ul{\t P}\\ &=
m^2_{\t A}\(\frac{4m_{\u A}}{\a_sm\dr{pl}}\)^n
\(\(\frac n{n+2}\)\frac{|\ul{\t P}|}{\t R(\t\G)}\)^\frac n2\(\frac{n-2}{n+2}\)
|\ul{\t P}|>0. \eal
\eqn26
$$
And eventually one gets
$$
z=z_0e^{\sqrt{\u A}\la}+z_0'e^{-\sqrt{\u A}\la}. \eqn27
$$
Taking $z_0'=0$ we get
$$
e^{2v}=z_0^2e^{2\sqrt{\u A}\la}. \eqn28
$$
Simply rescaling a time in the metric (14) we finally get
$$
ds^2_5=e^{2\sqrt{\u A}\la}\,dt^2 - d{\vec r}^{\,2} - r^2\,d\la^2. \eqn29
$$

In this way we get a funny toy model. We are confined on 3-dimensional
brane in a 4-dimensional euclidean space for a $\la=0$. Moreover $\la$ is
changing from~0 to~$2\pi$ resulting in some interesting possibilities of
communication and travel in extra dimensions. This is due to a fact that an 
\ef\ speed of light depends on~$\la$ in an \ex\ way:
$$
c\dr{eff}=e^{\sqrt{\u A}\la}. \eqn30
$$
If we move in $\la$ direction (even a little, remember $r=\frac1{m_{\u A}}$)
from~0 to $\la_0<2\pi$ and after this move in space direction from $\vec r_0$
to~$\vec r_1$ and again from~$\la_0$ to~0 (we \ef ly travel from $\vec r_0$
to~$\vec r_1$), then we can be in a point~$\vec r_1$ even very distant
from~$\vec r_0$ ($L=|\vec r_1-\vec r_0|$) in a much shorter time than~$\frac
Lc$ (where $c$ is the velocity of light taken to be equal~1). In some sense
we travel in hyperspace from {\it Star Wars} or {\it Wing Commander}.

Thus we consider a metric (see Ref. [2])
$$
ds^2_5=e^{2\sqrt{\u A}\la}\,dt^2 - d{\vec r}^{\,2} - r^2\,d\la^2. \eqn31
$$
We get the following geodetic equations for a signal travelling in~$\vec r$,
$\la$~direction with initial velocity at $\la=0$, $\frac{d\la}{dt}(0)=u$,
$\frac{d\vec r}{dt}(0)=\vec v$,
$$
\ealn{
\frac{d\vec r}{dt}&=\vec v\,e^{\sqrt{\u A}\la} &(32)\cr
\(\frac{d\la}{dt}\)^2&=e^{\sqrt{\u A}\la}-(1-u^2)e^{2\sqrt{\u A}\la}.&(33)
}
$$
One integrates
$$
\vec r(\la)=\vec r_0-\vec v_0\cdot \frac{2(1-u^2)^{3/2}}{\sqrt{\u A}}
\arctg\(\sqrt{1-\frac1{(1-u^2)^{1/2}}e^{-\sqrt{\u A}\la}}\), \eqn34
$$
$$
\la=\frac1{\sqrt{\u A}}\ln\(\frac{4(1-u^2)^{5/2}}{\u A(t-t_0)^2+4(1-u)^3}\)
\eqn35
$$
and
$$
\vec r(t)=\vec r_0 - \vec v_0\frac{(1-u^2)^{3/2}}{\sqrt{\u A}}\arctg
\(\sqrt{1+\frac{4(1-u^2)^2}{\u A(t-t_0)^2+4(1-u^2)^3}}\), \eqn36
$$
$$
0<\la<\min\[2\pi,\frac1{2\sqrt{\u A}}\ln\(\frac1{1-u^2}\)\]. \eqn37
$$

Let us consider a signal travelling along an axis $x$ from 0 to~$x_0$. The
time of this travel is simply~$x_0$. But now we can consider a travel from
zero to~$\la_0$ (in $\la$ direction) and after this from $(0,\la_0)$
to~$(x_0,\la_0)$ and after to~$(x_0,0)$. The time of this travel is as
follows: 
$$
t=2t_1+t_2 \eqn38
$$
where
$$
t_1=r \intop_0^{\la_0}\frac{d\la}{e^{\sqrt{\u A}\la}}
=\frac r{\sqrt{\u A}}\(1-e^{-\sqrt{\u A}\la_0}\) \eqn39
$$
and
$$
t_2=x_0e^{-\sqrt{\u A}\la_0}. \eqn40
$$
For $r$ is of order of a scale of G.U.T. the first term is negligible and
$$
t=x_0e^{-\sqrt{\u A}\la_0}. \eqn41
$$
Taking maximal value of $\la_0=2\pi$ we get
$$
t=x_0e^{-2\pi\sqrt{\u A}}. \eqn42
$$
Thus we achieve really shorter time of a travelling signal through the fifth 
dimension.

Let us consider the more general situation when we are travelling along a
time-like curve via fifth dimension from $\vec r=\vec r_0$ to $\vec r=\vec
r_1$. Let the parametric equations of the curve have the following shape:
$$
\displaylines{
\hphantom{(44)}\hfill\al
\vec r&=\vec r(\xi)\\
t&=t(\xi)\\
\la&=\la(\xi).
\eal
\hfill (43)\cr
\noalign{\vskip2pt}
0\le \xi\le 1,\cr
\noalign{\vskip2pt}
\la(0)=0, \ \la(1)=\la_0,\cr
\noalign{\vskip2pt}
\hphantom{(44)}\hfill\al
\vec r(0)&=\vec r_0\\
\vec r(1)&=\vec r_1.
\eal
\hfill (44)
}
$$
Let a tangent vector to the curve be $(\dot t(\xi),\dot {\vec r}(\xi),
\dot \la(\xi))$, where dot ``$\dot{\phantom t}$'' means a derivative \wrt
$\xi$. For a total time measured by a local clock of an observer one gets 
$$
\D t=\intop_0^1 d\xi\(e^{2\sqrt{\u A}\la(\xi)}
\(\frac{dt}{d\xi}\)^2 - \(\frac{d\vec r}{d\xi}\)^2 - r^2\(\frac{d\la}{d\xi}\)^2
\)^\frac12. \eqn45
$$
Let us consider a three segment curve such that
\item{1)} a straight line from $\vec r=0$, $\la=0$ to $\vec r=0$, $\la=\la_0$,
\item{2)} a straight line from $\vec r=0$, $\la=\la_0$ to $\vec r=\vec r_0$, $\la=\la_0$,
\item{3)} a straight line from $\vec r=\vec r_0$, $\la=\la_0$ to $\vec r=\vec
r_0$, $\la=0$.

The additional assumptions are such that the traveller travels with a
constant velocity $v_\la$ on the first segment and on the third segment and
with a velocity $\vec v$ on the second one.

Thus we have a parametrization:

1) On the first segment
$$
\al
\vec r&=0\\
t&=\frac{\la_0}{v_\la}\,\xi, \quad 0\le \xi\le 1\\
\la&=\xi\la_0.
\eal
\eqn46
$$

2) On the second segment
$$
\al
\vec r(\xi)&=\vec r_0\xi\\
t&=\frac{|\vec r_0|}{|\vec v|}\,\xi, \quad 0\le \xi\le 1\\
\la&=\la_0.
\eal
\eqn47
$$

3) On the third segment
$$
\al
\vec r&=\vec r_0\\
t&=\frac{\la_0}{|\vec v|}\,(1-\xi), \quad 0\le \xi\le 1\\
\la&=(1-\xi)\la_0.
\eal
\eqn48
$$

From Eq.\ (45) one easily gets
$$
\ealn{
\D t&=\D t_1 + \D t_2 &(49)\cr
\D t_1&=\frac{2r}{\sqrt{\u A}}\Biggl(
\frac1{\sqrt{v_\la \la_0}}\(\sqrt{e^{2\sqrt{\u A}\la_0}-v_\la \la_0}
-\sqrt{1-v_\la \la_0}\)\cr
&+\arctg\Biggl(\sqrt{\frac{e^{2\sqrt{\u A}\la_0}-v_\la \la_0}{v_\la \la_0}}\Biggr)
-\arctg\(\sqrt{\frac{1-v_\la \la_0}{v_\la \la_0}}\)\Biggr)&(50)\cr
\D t_2&=\frac{|\vec r_0|}{|\vec v|} \sqrt{e^{2\sqrt{\u A}\la_0}-|\vec v|^2}
&(51)
}
$$

For $r$ is of order of an inverse scale of G.U.T., the first term in Eq.\
(49) is negligible and
$$
\D t \simeq \frac{|\vec r_0|}{|\vec v|} \sqrt{e^{2\sqrt{\u A}\la_0} -|\vec
v|^2}. \eqn52a
$$
The traveller passing from $\vec r=0$ to $\vec r=\vec r_0$ (having $\la=0$
during his travel) reached $\vec r=\vec r_0$ at time
$$
\D t'=\frac{|\vec r_0|}{|\vec v|} \(1 -|\vec v|^2\)^{1/2}. \eqn52b
$$
The first traveller has the following condition for $\vec v$:
$$
|\vec v|\le e^{\sqrt{\u A}\la_0}, \quad 0\le \la_0\le 2\pi. \eqn53a
$$
The second
$$
|\vec v|\le 1. \eqn53b
$$
Thus we see that we can travel quicker taking a bigger $\vec v$.

For an unmoving observer the time needed for a travel in the first case is
$$
\al
&\D t=\frac{2r\la_0}{v_\la}+\frac{|\vec r_0|}{|\vec v|}\simeq \frac{|\vec
r_0|} {|\vec v|},\\
&|\vec v|\le e^{\sqrt{\u a}\la_0}, \quad 0\le \la_0\le 2\pi.
\eal
\eqn52a$*$
$$
In the second case
$$
\al
\D t'&=\frac{|\vec r_0|}{|\vec v|},\\
|\vec v|&\le 1
\eal
\eqn52b$*$
$$
Thus in some sense we get a dilatation of time:
$$
\D t'=\(e^{2\sqrt{\u A}\la_0}-{\vec v}^{\,2}\)^{1/2}\D t \eqn52a$**$
$$
in the first case,
$$
\D t'=(1-{\vec v}^{\,2})^{1/2}\D t \eqn52b$**$
$$
in the second case.

Let us come back to the Eqs (35--37). We see that a particle coming along
a geodesic is confined in a shell surrounding a brane in the fifth dimension
$$
0\le \la\le \frac r{\sqrt{\u A}}\ln\(\frac1{(1-u^2)^{1/2}}\). \eqn54
$$
The particle oscillates in the shell. Moreover $0\le \la \le 2\pi$, thus if
$\la$ excesses $2\pi$, we take for a new value $\la-2\pi$ (i.e.\
modulo~$2\pi$). Usually for~$u$ not close to~1 this is a really thin shell
of order $r$ (a~length of a unification scale).

Let us come back to our toy model for a cone with $\t A=0$. In this case an
equation of cone in 5-dimensional Minkowski space is
$$
x^2+y^2+z^2+r^2\la^2=t^2. \eqn55
$$
An interesting question is what is an analog for (55) if $\u A\ne 0$. One
easily gets that
$$
x^2+y^2+z^2+r^2\la^2=t^2r^2\la^2{\t A}^2e^{2\sqrt{\u A}\la}. \eqn56
$$

There is no simple way to get Eq.\ (55) from Eq.\ (56). Moreover if
we change a coordinate~$t$ into $rt\la\t A$ one gets
$$
x^2+y^2+z^2+r^2\la^2=t^2e^{2\sqrt{\u A}\la}. \eqn57
$$
and for $\t A=0$ we get Eq. (55).

Let us notice that Eq.\ (57) is necessary to be considered for $\la$
modulo~$2\pi$. It means even $\la$ could be any real number in the equation
we should put in a place of~$\la$, $[\la]$, where $[\la]$ is defined as
follows 
$$
\la=[\la]+2\pi n, \quad [\la]\in \langle0,2\pi),
$$
$n$ is an integer and $\la\in(-\infty,+\infty)$. In this way in a place of
Eq.~(57) we get
$$
x^2+y^2+z^2+r^2[\la]^2=t^2e^{2\sqrt{\u A}[\la]}. \eqn57$*$
$$
The last equation defines an interesting 4-dimensional hypersurface in
5-dimensional space (${\Bbb R}^5$).

For $\la=0$ we get a light cone on a brane (4-dimensional Minkowski space).
For branes with $0<\la=\la_0<2\pi$ we have a ``cone''
$$
x^2+y^2+z^2=t^2e^{2\sqrt{\u A}\la_0}-\la_0^2 \eqn58
$$
which is a 3-dimensional hyperboloid.

In our model we get in a quite natural way a warp factor known from
Ref.~[3]. Moreover we have a different reason to get it. The reason of this
warp factor is a many dimensional dependence of a scalar field
$\varPsi=\varPsi(x,\la)$. In particular higher-dimensional excitations of~$\varPsi$
forming a tower of massive scalar fields. In this model $\varPsi$ can be
developed in a Fourier series
$$
\varPsi(x,\la)=\sum_{m=0}^\infty \psi^1_m(x)\cos(m\la)+
\sum_{m=1}^\infty \psi^2_m(x)\sin(m\la) \eqn59
$$
where
$$
\ealn{
\psi^1_m(x)&=\frac1{2\pi}\intop_0^{2\pi}\varPsi(x,\la)\cos(m\la)\,d\la
&(60)\cr 
\psi^2_m(x)&=\frac1{2\pi}\intop_0^{2\pi}\varPsi(x,\la)\sin(m\la)\,d\la
&(61)\cr 
&m=1,2,\ldots\cr
\frac{\psi^1_0(x)}2&=\frac1{2\pi}\intop_0^{2\pi}\varPsi(x,\la)\,d\la. &(62)
}
$$

Thus an interesting point will be to find a spectrum of mass for this tower. 
This is simply
$$
m_m^2=m_{\u A}^2 \(\o M+\frac{n^2\zeta^2}{\zeta^2+1}\)m^2, \qquad
m=0,1,2,\ldots \eqn63
$$
Now we come to the calculation of a content of a warp factor (a content in
terms of a tower of scalar fields). Thus we calculate
$$
\ealn{
\psi^1_m&=\frac1{2\pi}\intop_0^{2\pi} e^{2\sqrt{\u A}\la}\sin(m\la)\,d\la,
\qquad m=1,2,3,\ldots &(64)\cr
\psi^2_m&=\frac1{2\pi}\intop_0^{2\pi} e^{2\sqrt{\u A}\la}\cos(m\la)\,d\la,
\qquad m=1,2,3,\ldots &(65)\cr
\frac12 \psi^1_0&=\frac1{2\pi}\intop_0^{2\pi} e^{2\sqrt{\u A}\la}\,d\la.
&(66)
}
$$
One gets
$$
\ealn{
\frac12\psi^1_0&=\frac1{4\pi\sqrt{\t A}}\(e^{4\pi\sqrt{\u A}}-1\) &(67)\cr
\psi^1_m&=-\frac m{2\pi}\frac{(e^{4\pi\sqrt{\u A}}-1)}{(4\t A+m^2)}
&(68)\cr
\psi^2_m&=\frac{\sqrt{\t A}(e^{4\pi\sqrt{\u A}}-1)}{\pi(4\t A+m^2)}\,.
&(69)
}
$$
The simple question is what is the energy to excite the warp factor in terms of
a tower of scalar fields. The answer is as follows:
$$
E\dr{wf}=\intop_0^{2\pi} \br{scal}T_{\kern-5pt 44}\bigl(e^{2\sqrt{\u A}\la}
\bigr)\,d\la \eqn70
$$
where $\br{scal}T_{\kern-5pt 44}$ is a time component of an energy-momentum
tensor for a scalar field calculated for $\varPsi=e^{2\sqrt{\u A}\la}$.

One gets
$$
\ealn{
&\br{scal}T_{\kern-5pt 44}=\frac12 \(\frac{d\varPsi}{d\la}\)^2 + 
\frac12 \la_{c_0}(\varPsi),
\quad \la_{c_0}(\varPsi)=\frac12|\g|e^{n\varPsi}-\frac\b2 e^{(n+2)\varPsi} &(71)\cr
&\intop_0^{2\pi}\frac12 \(\frac d{d\la}e^{2\sqrt{\u A}\la}\)^2\,d\la =
\frac{\sqrt{\t A}}2\(e^{8\pi\sqrt{\u A}}-1\) &(72)
}
$$
and
$$
\frac12 \la_{c_0}\bigl(e^{2\sqrt{\u A}\la}\bigr)=
-\frac12\, e^{n(e^{2\sqrt{\u A}\la})}
\(\b e^{2(e^{2\sqrt{\u A}\la})}-|\g|\), \eqn73
$$
where
$$
\al
\g&=\frac{m^2_{\u A}}{\a_s^2}\ul{\t P}=\frac1{r^2}\ul{\t P}, \\
\b&=\frac{\a_s^2}{l\dr{pl}^2}\t R(\t\G)={\a_s^2}m\dr{pl}^2\t R(\t\G) .
\eal
$$

One easily gets
$$
\al
E\dr{wf}&=\frac{\sqrt{\t A}}2 \bigl(e^{8\pi\sqrt{\u A}}-1\bigr)
-\frac1{4\sqrt{\t A}} \Biggl[|\g|\Biggl(\Li
\Biggl(\exp\biggl(\frac{e^{4\pi\sqrt{\u A}}}n\biggr)\Biggr)
-\Li\bigl(\root n\of e\bigr)\Biggr)\\
&-\b\Biggl(\Li\Biggl(\exp\biggl(\frac{e^{4\pi\sqrt{\u A}}}{(n+2)}\biggr)\Biggr)-\Li\bigl(
\root{n+2}\of e\bigr)\Biggr)\Biggr]
\eal \eqn74
$$
where
$$
\Li(x)=\dint_0^x\frac{dz}{\ln z},\quad x>1, \eqn75
$$
is an integral logarithm. The integral is considered in the sense of a
principal value.

Let us consider Eq.\ (74) using asymptotic formula for $\Li(x)$ for large~$x$:
$$
\Li(x) \sim x\biggl[\frac1{\ln x}+\sum_{m=1}^\infty
\frac{m!}{\ln^{m+1}x}\biggr].
$$
One gets
$$
\al
E\dr{wf}&=\frac{\sqrt{\t A}}2 \(e^{8\pi\sqrt{\u A}}-1\)-
\frac{e^{-4\pi\sqrt{\u A}}}{4\sqrt{\t A}}\Biggl(n|\g|\exp
\biggr(\frac{e^{4\pi\sqrt{\u A}}}n\biggl)\\
&-(n+2)\b\exp\biggl(\frac{e^{4\pi\sqrt{\u A}}}{n+2}\biggr)\Biggr).
\eal \eqn74a
$$

However, let us think quantum-mechanically in a following direction. Let us
create (maybe in an accelerator as some kind of resonances) scalar particles
of mass $m_n$ (Eq.~(63)) with an amount of $|\psi^1_m|^2+
|\psi^2_m|^2$. In this way we create the full warp factor (for they due to
interference will create it as a Fourier series). The question is what is an
energy needed to create those particles. The answer is simply
$$
\o E=\sum_{m=1}^\infty \(|\psi^1_m|^2+|\psi^2_m|^2\)m_m. \eqn76
$$
One easily gets
$$
\o E=\frac{m_{\u A}}{4\pi^2}\(e^{4\pi\sqrt{\u A}}-1\)^2\(\o M+\frac{n^2\zeta^2}
{\zeta^2+1}\)^\frac12 \sum_{m=1}^\infty \frac m{4\t A+m^2}\,.\eqn77
$$
The series $\sum_{m=1}^\infty \frac m{4\u A+m^2}$ is divergent for
$$
\frac m{4\t A+m^2}\simeq \frac 1m\,. \eqn78
$$
Thus we need a regularization technique to sum it. We use $\zeta$-function
regularization technique similar as in Casimir-effect theory. Let us
introduce a $\zeta_{\u A}$-function on a complex plane
$$
\zeta_{\u A}=\sum_{m=1}^\infty \(\frac m{4\t A+m^2}\)^s \eqn79
$$
for $\Re(s)\ge1+\d$, $\d>0$.

This function can be extended analytically on a whole complex plane.
Obviously $\zeta_{\u A}$ has a pole for $s=1$ (as a Riemann $\zeta$
function). Thus we should regularize it at $s=1$. One gets
$$
\o E=\frac{m_{\u A}}{4\pi^2}\(e^{4\pi\sqrt{\u A}}-1\)^2\(\o M+\frac{n^2\zeta^2}
{\zeta^2+1}\)^\frac12 \zeta^r_{\u A}(1)\eqn80
$$
where $\zeta^r_{\u A}$ means a regularized $\zeta_{\u A}$.

Let us introduce a $\zeta$-Riemann function
$$
\zeta\dr{R}(s)=\sum_{m=1}^\infty \frac1{m^s},
\qquad \Re(s)\ge1+\d,\ \d>0. \eqn81
$$

One can write
$$
\hbox{``}\zeta_{\u A}(1)\hbox{''}=
\hbox{``}\zeta_{\text{R}}(1)\hbox{''}+4\t A\sum_{m=1}^\infty \frac1{m(4\t A+m^2)}
\eqn82
$$
or
$$
\hbox{``}\zeta_{\u A}(1)\hbox{''}=
\hbox{``}\zeta_{\text{R}}(1)\hbox{''}+4\t A\zeta\dr{R}(3)
-16{\t A}^2\sum_{m=1}^\infty \frac1{m^3(4{\t A}^2+m^2)}\,. \eqn83
$$
The series in Eqs (82--83) are convergent. Moreover $\zeta\dr R(s)$ has
a pole at $s=1$ and
$$
\lim_{s\to1}\[\zeta\dr R(s)-\frac1{s-1}\]=\g_E \eqn84
$$
where $\g_E$ is an Euler constant.

Thus we can regularize $\zeta_{\u A}(s)$ in such a way that
$$
\zeta^r_{\u A}(s)=\zeta\dr{R}(s)-\frac1{s-1}+4\t A\sum_{m=1}^\infty
\frac1{m(4\t A+m^2)}\,. \eqn85
$$
Thus
$$
\zeta^r_{\u A}(1)=\g_E + 4\t A\sum_{m=1}^\infty \frac1{m(4\t A+m^2)}\,.
\eqn86
$$
The series
$$
\sum_{m=1}^\infty \frac1{m(4\t A+m^2)} \eqn87
$$
is convergent for every $z^2=-4\t A$ and defines an analytic function
$$
f(z)=\sum_{m=1}^\infty \frac1{m(m^2-z^2)}\,. \eqn88
$$
Using some properties of an expansion in simple fraction for
$$
\psi(x)=\frac d{dx}\log \G(x) \eqn89
$$
where $\G(x)$ is an Euler $\G$-function, one gets
$$
\ealn{
\psi(x)&=-\g_E - \sum_{m=0}^\infty \(\frac1{m+x}-\frac1{m+1}\)&(90)\cr
\zeta^r_{\u A}(1)&=-\frac12\(\psi(2\sqrt{\t A})+\psi(-2\sqrt{\t A})\)
&(91)
}
$$
or
$$
\zeta^r_{\u A}(1)=\frac1{4\sqrt{\t A}}+\frac\pi2 \ctg2\sqrt{\t A}\pi \eqn92
$$
(where we use $\G(z)\G(-z)=-\frac \pi{z\sin\pi z}$).

And finally
$$
\o E=\frac{m_{\u A}}{8\pi^2}\(\o M+\frac{n^2\zeta^2}{\zeta^2+1}\)
\(e^{4\pi\sqrt{\u A}}-1\)^2\(\frac1{2\sqrt{\t A}}+\pi\ctg(2\sqrt{\t A}\pi)\).
\eqn93
$$
It is easy to see that $\o E=0$ for $\t A=0$,
$$
\sqrt{\t A}=m_{\u A}\(\frac{4\sqrt n\, m_{\u A}}{\a_s\sqrt{n+2}\,m\dr{pl}}\)^\frac
n2 \(\frac{|\ul{\t P}|}{\t R(\t \G)}\)^\frac n4
\(\frac{n-2}{n+2}\,|\ul{\t P}|\)^\frac12. \eqn94
$$
For we use a dimensionless coordinate $\la$ (not $\t \la=r\cdot \la=\frac
\la{m_{\u A}}$), we can omit a factor $m_{\u A}$ in front of the right-hand
side in the formula (94).

It is interesting to notice that for $\t A=0$, $e^{2\sqrt{\u A}}=1$ and we
have to do with a Fourier expansion of a constant mode only ($m=0$).
This mode is a massless mode. However, a massless mode can obtain a mass from
some different mechanism. Thus in some sense we have to do
with an energy
$$
\frac14 \Bigl|\psi^1_0(\t A=0)\Bigr|^2 m_\varPsi=m_\varPsi\,.\eqn95
$$

One gets the following formula
$$
m_\psi=m\dr{pl} \sqrt{\frac{|\ul{\t P}|(n+2)}{(\o M+\frac{n^2\zeta^2}{\zeta^2+1})}}
\(\frac{4\sqrt n\,\sqrt{|\ul{\t P}|}m_{\u A}}{\a_s\sqrt{n+2}\,\sqrt{\t R(\t\G)}\,
m\dr{pl}}\)^\frac n2.\eqn96
$$
Thus it seems that in order to create a warp factor $e^{2\sqrt{\u A}\la}$ in
a front of $dt^2$ we should deposite an energy of a tower of scalar particles
$$
\t E=E+m_\psi\,.\eqn97
$$
Even a factor equal to 1 is not free. This is of course supported by a
classical field formula (Eq.~(75)). For $\t A=0$ we get
$$
E\dr{wf}=-\pi\cdot e^n(\b e^2-|\g|) \eqn98
$$
or
$$
E\dr{wf}=-\frac{\pi e^n}{\a_s^2}\(e^2\a_s^4m\dr{pl}^2 \t R(\t \G)-m^2_{\u A}
|\ul{\t P}|\). \eqn99
$$

To be honest we should multiply $E\dr{wf}$ and $E$ by $V_3$, where $V_3$ is a
volume of a space and a result will be divergent. Moreover, if we want to
excite a warp factor only locally, $V_3$~can be finite and the result also
finite. 

Let us come back to the Eq.\ (4). The variational principle based on
(4) \lg\ is really in $(n_1+4)$-dimensional space $E\times G/G_0$, where
a size of a compact space $M=G/G_0$ is~$r$. In this way the gravity
lives on $(n_1+4)$-dimensional manifold, where $n_1$ dimensions are curled
into a compact space. The scalar field $\varPsi$ lives also on
$(n_1+4)$-dimensional manifold. The matter is 4-dimensional. Thus we can
repeat some conclusions from Randall-Sundrum ([4]) and Arkani-Hamed, Savas
Dimopoulos and Dvali ([5]) theory. Similarly as in their case we have
gravity for two regions ($V(R)$ is a Newtonian \gr\ potential).

1) $R\ll r$
$$
V(R)\cong \frac{m_1\cdot m_2}{m^{n_1+2}\dr{pl$(n_1+4)$}}\,
\frac1{R^{n_1+1}}\,, \eqn100
$$

2) $R\gg r$
$$
V(R)\cong \frac{m_1\cdot m_2}{m^{n_1+2}_{\text{pl}(n_1+4)}r^{n_1}}\,
\frac1{R}\,, \eqn101
$$
where $m\dr{pl$(n_1+4)$}$ is a $(n_1+4)$-dimensional Planck's mass. So
$$
m^2\dr{pl}=m^{n_1+2}_{\text{pl}(n_1+4)}\cdot r^{n_1}=
\frac{m^{n_1+2}_{\text{pl}(n_1+4)}}{m^{n_1}_{\u A}}=
\(\frac{m_{\text{pl}(n_1+4)}}{m_{\u A}}\)^{n_1}\cdot m^2_{\text{pl}
(n_1+4)}\,.  \eqn102
$$
Thus
$$
m_{\text{pl}(n_1+4)}=(m\dr{pl})^\frac2{n_1+2}\cdot (m_{\u A})^\frac
{n_1}{n_1+2}. \eqn103
$$

Eqs (102--103) give us a constraint on parameters in our theory and
establish a real strength of \gr\ interactions given by
$m_{\text{pl}(n_1+4)}$. 

However, in our case we have to do with a scalar-tensor theory of gravity.
Thus our \gr\ \ct\ is \ef
$$
G\dr{eff}=G_Ne^{-(n+2)\varPsi}=G_N\rho^{n+2}. \eqn104
$$
In this case we have to do with \ef\ ``Planck's masses'' in $n_1+4$ and
4~dimensions 
$$
\ealn{
m^2_{\text{pl}(n_1+4)}&\longrightarrow m^2_{\text{pl}(n_1+4)}
\cdot e^{(n+2)\varPsi_{(n_1+4)}} &(105)\cr
m^2_{\text{pl}}&\longrightarrow m^2_{\text{pl}}
\cdot e^{(n+2)\varPsi} &(106)
}
$$
where $\varPsi_{(n_1+4)}$ is a scalar field $\varPsi$ for $R\ll r$ and $\varPsi$ is
for $R\gg r$.

In this way Eq.\ (102) reads
$$
m^2\dr{pl} e^{(n+2)\varPsi}=
\(\frac{m_{\text{pl}(n_1+4)}e^{\frac{(n+2)}2 \varPsi_{(n_1+4)}}}
{m_{\u A}}\)^{n_1}\cdot
m^2_{\text{pl}(n_1+4)}\cdot e^{(n+2)\varPsi_{(n_1+4)}}. \eqn107
$$
For our contemporary epoch we can take $e^{(n+2)\varPsi}\simeq1$.

If we take as in the case of Ref.\ [5]
$$
m_{\text{pl}(n_1+4)}\simeq m\dr{EW}, \eqn108
$$
where $m\dr{EW}$ is an electro-weak energy scale. We get
$$
\(\frac{m\dr{pl}}{m\dr{EW}}\)^2 = \(\frac{m\dr{EW}}{m_{\u A}}\)^{n_1}
\cdot e^{\frac12 (n+2)(n_1+2)\varPsi_{(n_1+4)}}. \eqn109
$$
For $R\ll r$ the field $\varPsi_{(n_1+4)}$ has the following behaviour
$$
\varPsi_{(n_1+4)}\cong \varPsi_0 +\frac \a{R^{n_1+1}}\,. \eqn110
$$

Thus we can approximate $\varPsi_{(n_1+4)}$ by $\varPsi_0$ where $\varPsi_0$ is a
critical point of a selfinteracting potential for~$\varPsi$. In this way one gets
$$
\(\frac{m\dr{pl}}{m\dr{EW}}\)\(\frac{m_{\u A}}{m\dr{EW}}\)^{n_1}
=\exp\(\frac{(n+2)(n_1+2)}2\,\varPsi_0\) \eqn111
$$
where 
$$
e^{\varPsi_0}=\sqrt{\frac{n|\g|}{(n+2)\b}}=
\frac1{\a_s}\(\frac{m_{\u A}}{m\dr{pl}}\)\sqrt{\frac{n\ul{\t P}}
{(n+2)\t R(\t\G)}}. 
$$
Thus one gets:
$$
\al
&\(\frac{m\dr{pl}}{m\dr{EW}}\)\(\frac{m_{\u A}}{m\dr{EW}}\)^{n_1}
\(\frac{m\dr{pl}}{m_{\u A}}\)^\frac{(n+2)(n_1+2)}2\\
&\qquad{}=
\(\frac1{\a_s}\)^\frac{(n+2)(n_1+2)}2
\(\frac{n|\ul{\t P}|}{(n+2)|\t R(\t\G)|}\)^\frac{(n+2)(n_1+2)}4. 
\eal
\eqn112
$$
It is easy to see that we can achieve a solution for a hierarchy problem if
$\biggl|\dfrac{\ul{\t P}}{\t R(\t \G)}\biggr|$ is sufficiently big, e.g. taking $\t R(\t\G)$
sufficiently small and/or $\ul{\t P}$ sufficiently big playing with \ct s $\xi$
and~$\zeta$. In this way a real strength of gravity will be the same as
electro-weak interactions and the hierarchy problem has been reduced to the
problem of smallness of a \co\ \ct\ (i.e.\ $\t R(\t\G)\simeq0$).

Let us consider Eq.\ (93) in order to find such a value of~$\t A$ for
which $\o E$ is minimal. One finds that
$$
\o E=0 \eqn113
$$
for
$$
\t A=\frac{x^2}{4\pi^2} \eqn114
$$
where $x$ satisfies an equation
$$
x=-\tg x,\quad x>0. \eqn115
$$
Eq.\ (115) has an infinite number of roots. One can find some roots of
Eq.~(115) for $x>0$. We get
$$
\al
x_1&=4.913\dots\\
x_2&=7.978\dots\\
x_3&=11.086\dots\\
x_4&=14.207\dots\\
\eal \eqn116
$$
We are interested in large roots (if we want to have $c\dr{eff}$ considerably
big). 

In this case one finds
$$
x=\e+\tfrac\pi2(2l+1) \eqn117
$$
where $\e>0$ is small and $l=1,2,\ldots$. Moreover, to be in line in our
approximation, $l$~should be large. One gets
$$
\tfrac\pi2(2l+1)+\e=\ctg\e. \eqn118
$$
For $\e$ is small, one gets
$$
\ctg\e \simeq \frac1\e\,. \eqn119
$$

Finally one gets 
$$
\e^2+(2l+1)\tfrac\pi2-1=0 \eqn120
$$
and
$$
\ealn{
\e&\simeq \frac2{(2l+1)\pi} &(121)\cr
x&=\frac\pi2\,(2l+1)+\frac2{(2l+1)\pi} &\text{(121a)}\cr
\t A&=\frac1{4\pi^2}\(\frac\pi2 \,(2l+1)+\frac2{(2l+1)\pi}\)^2.&(122)
}
$$
It is easy to notice that our approximation (121a) is very good even for
small~$l$, e.g.\ for $l=4$ we get $x_4$ (Eq.~(116)).

In this way an \ef\ velocity of light can be arbitrarily large,
$$
c\dr{eff}\gi{max}=c\exp\(\frac12\(\frac\pi2(2l+1)+\frac2{(2l+1)\pi}\)\)\eqn123
$$
(i.e.\ for $\la=2\pi$). However, in order to get such large $c\dr{eff}$ we
should match Eq.~(94) with (122). One gets
$$
\frac\pi2\,(2l+1)+\frac2{(2l+1)\pi}
=2\pi\(\frac{4\sqrt n\, m_{\u A}}{\a_s\sqrt{n+2}\,m\dr{pl}}\)^\frac
n2 \(\frac{|\ul{\t P}|}{\t R(\t \G)}\)^\frac n4
\(\frac{n-2}{n+2}\,|\ul{\t P}|\)^\frac12. \eqn124
$$

In this way we get some interesting relations between parameters in our
theory. To have a large $c\dr{eff}$ we need to make $|\ul{\t P}|$ large and
$\t R(\t \G)$ small. In this case only the first term of the left-hand side
of (124) is important and we should play with the ratio $\(\dfrac
{m_{\u A}}{\a_sm\dr{pl}}\)$ in order to get the relation
$$
(2l+1)
=4\(\frac{4\sqrt n\, m_{\u A}}{\a_s\sqrt{n+2}\,m\dr{pl}}\)^\frac
n2 \(\frac{|\ul{\t P}|}{\t R(\t \G)}\)^\frac n4
\(\frac{n-2}{n+2}\,|\ul{\t P}|\)^\frac12 \eqn125
$$
where $l$ is a big natural number. In this way we can travel in hyperspace
(through extra dimension) almost for free. The energy to excite the warp factor
is zero (or almost zero). Thus it could be created even from a fluctuation in
our Universe.

Let us notice that Eq.\ (115) is in some sense universal for a warp factor.
It means that it is the same for any realization of the Nonsymmetric
Jordan-Thiry (Kaluza-Klein) Theory.

However, if we take Eq.\ (74) and if we demand
$$
E\dr{wf}=0, \eqn126
$$
we get some roots (they exist) depending on $n$, $\b$ and $|\g|$. Thus the
roots, some values of $\t A>0$ depend on details of the theory, they are not
universal as in the case of Eqs~(113--115). In this case the Eq.~(94) is
in some sense a selfconsistency condition for a theory and more restrictive.
The fluctuations of a tower of scalar fields in the case of $\sqrt{\t
A}=\frac14(2l+1)$ are as follows
$$
\ealn{
\frac12\psi^1_0 &=\frac{\(e^{\pi(2l+1)}-1\)}{2l+1} &(127)\cr
\psi^1_m&=-\frac m\pi \cdot \frac{\(e^{\pi(2l+1)}-1\)}{\((2l+1)^2+4m^2\)}
&(128)\cr
\psi^2_m&=\frac{(2l+1)\(e^{\pi(2l+1)}-1\)}{\pi\((2l+1)^2+4m^2\)}\,. &(129)
}
$$
In the case of very large $l$ (which is really considered) one gets
$$
\ealn{
\frac12\psi^1_0&=\frac{e^{\pi(2l+1)}}{2l} &\text{(127a)}\cr
\psi^1_m&=-\frac m{2\pi}\cdot \frac{e^{\pi(2l+1)}}{4(l^2+m^2)}
&\text{(128a)}\cr 
\psi^2_m&=\frac{l\cdot e^{\pi(2l+1)}}{2\pi(l^2+m^2)} &\text{(129a)}\cr
&m=1,2,\ldots
}
$$
According to our calculations the energy of the fluctuations (excitations)
(127a--129a) is zero.

Thus the only difficult problem to get a warp factor is to excite a tower of
scalar fields coherently in a shape of (127--129a) or to wait for such a
fluctuation. In the first case it is possible to use some techniques from
quantum optics applied to the scalar fields~$\psi_m$. In our case we get
$$
\ealn{
\psi_0&=\frac{e^{\pi(2l+1)}}{2l} &(130)\cr
\psi_m&=\frac{e^{\pi(2l+1)}\sqrt{16l^2+m^2}}{8\pi(l^2+m^2)} \sin(m\la+\d_m)
e^{im_mt} &(131)
}
$$
where
$$
\tg \d_m=-\frac m{4l} \eqn132
$$
and $m_m$ is given by the formula (63).

For $\psi_0$ is \ct\ in time and does not contribute to the total energy of a
warp factor (only to the energy of a warp factor equal to~1), we consider
only 
$$
\psi(\la,t)=\frac{e^{\pi(2l+1)}}{8\pi}
\biggl[\sum_{m=1}^\infty \frac{\sqrt{16l^2+m^2}}{l^2+m^2}
\sin(m\la+\d_m)e^{im_mt}+\frac{4\pi}{l}\biggr].
\eqn133
$$
For $t=0$ we get
$$
\psi(\la,0)=e^{\pi(2l+1)\la} \eqn134
$$
and the wave packet (133) will disperse. Thus we should keep the wave
packet to not decohere in such a way that
$$
\psi(\la,t)=\frac{e^{\pi(2l+1)}}{8\pi}
\biggl[\sum_{m=1}^\infty \frac{\sqrt{16l^2+m^2}}{l^2+m^2}
\sin(m\la+\d_m)+\frac{4\pi}{l}\biggr].
\eqn135
$$
It means $\psi(\la,t)$ should be a pure zero mode for all the time $t\ge0$.

Let us come back to the Eq.\ (125) (earlier to Eq.~(94)) taking under
consideration that in a real model $m_{\u A}=\frac{\a_s}r$). One can rewrite
Eq.~(125) in the following way:
$$
\(\frac{2l+1}{4\a_s}\)^{4/n}\(\frac{n+2}{(n-2)|\uP|}\)^{2/n}
\(\frac{\a_s\sqrt{n+2}\,m\dr{pl}}{4\sqrt n\, m_{\u A}}\)^2= \frac{|\uP|}
{\RG} \eqn136
$$
or
$$
\RG=\(\frac{4\a_s}{2l+1}\)^{4/n}\(\frac{(n-2)|\uP|}{n+2}\)^{2/n}
\(\frac{4\sqrt n\, m_{\u A}}{\a_s\sqrt{n+2}\,m\dr{pl}}\)^2 |\uP|=
2F(\z). \eqn137
$$

We want to find some conditions for~$\mu$ and~$\z$ (or~$\xi$ and~$\z$) in
order to satisfy Eq.~(94) for large~$l$ in a special model with $H=G2$,
$\dim G2=n=14$ for $M=S^2$ and $\RG$ calculated for $G=SO(3)$. In this case
one finds
$$
F(\z)=3^{2/7}\cdot 7 \cdot \(\frac{\a_s}{l}\)^{4/7} \(\frac{m_{\u A}}
{\a_s m\dr{pl}}\)^2 |\uP|^{15/7} \eqn138
$$
where $\uP$ is given by the formula 
$$
\al
\uP (\zeta ) &=\biggl\{{16|\zeta |^3(\zeta ^2+1)\over 3(2\zeta
^2+1)(1+\zeta ^2)^{5/2}}\(\zeta ^2E\({|\zeta |\over
\sqrt{\zeta ^2+1}}\biggr)-(2\zeta ^2+1)K\biggl({|\zeta |\over
\sqrt{\zeta ^2+1}}\)\)\cr
&+ 8\ln (|\zeta |\sqrt{\zeta ^2+1}) + {4(1+9\zeta ^2-8\zeta ^4)|\zeta
|^3\over 3(1+\zeta ^2)^{3/2}}\biggr\}
\biggm{/} (\ln(|\zeta |+\sqrt{\zeta ^2+1})+2\zeta ^2+1).
\eal
$$
 
We calculate $\RG$ for a group $SO(3)$ in the first point of Ref.~[1]
and we get
$$
\RG=\frac{2(2\mu^3+7\mu^2+5\mu+20)}{(\mu^2+4)^2}\,. \eqn139
$$
The polynomial
$$
W(\mu)=2\mu^3+7\mu^2+5\mu+20 \eqn140
$$ 
possesses only one real root
$$
\mu_0=-\frac{\root3\of{1108+3\sqrt{135645}}}{6}-\frac76-\frac{19}
{6\cdot\root3\of{1108+3\sqrt{135645}}}=-3.581552661\ldots.\eqn141
$$ 
It is interesting to notice that $W(-3.581552661)=2.5\cdot10^{-9}$ and for
70-digit approximation of~$\mu_0$, $\t \mu$ equal to
$$
-3.581552661076733712599740215045436907383569800816123632201827285932446,
$$
we have
$$
W(\t\mu)=0.1\cdot 10^{-67}.
$$ 
The interesting point in $\t\mu$ is that any truncation of~$\t\mu$ (it means,
removing some digits from the end of this number) results in growing $W$,
i.e., 
$$
0<W(\t\mu_n)<W(\t\mu_{n-1}),
$$
where $\t\mu_n$ means a truncation of~$\t\mu$ with only $n$~digits, $n\le70$,
$\t\mu_{70}=\t\mu$. This can help us in some approximation for $\RG>0$,
close to zero.

Substituting (139) into Eq.\ (137) we arrive to the following equation:
$$
-F(\z)\mu^4+2\mu^3+\mu^2(7-4F(\z))+5\mu+4(-2F(\z)+5)=0. \eqn142
$$
Considering $F(\z)$ small we can write a root of this polynomial as
$$
\mu=\mu_0+\e \eqn143
$$
where $\mu_0$ is the root of the polynomial (140) and $\e$ is a small
correction. In this way one gets the solution
$$
\mu=-3.581552661\ldots +47\(\frac{\a_s}{l}\)^{4/7}\(\frac{m_{\u A}}
{m\dr{pl}}\)^2q(\z) \eqn144
$$
where $q(\z)$ is given by the formula
$$
\al
q(\z)&=\Biggl
| \frac{16|\z|^3(\z^2+1)}{3(2\z^2+1)(1+\z^2)^{5/2}}
\(\z^2E\(\frac{|\z|}{\sqrt{\z^2+1}}\)-(2\z^2+1)K\(\frac{|\z|}{\sqrt{\z^2+1}}\)\)\cr
&\qquad{}+8\ln\(|\z|\sqrt{\z^2+1}\)+\frac{4(1+9\z^2-8\z^4)|\z|^3}
{3(1+\z^2)^{3/2}}\Biggr|^\frac{15}7\cr
&\times\[\ln\(|\z|+\sqrt{\z^2+1}\)+2\z^2+1\]^{-\frac{15}7},
\eal
\eqn145
$$
$|\z|>|\z_0|=1.36$ (see Ref.\ [1]), and $K$ and $E$ are given by the formulae
$$
K(k) = \intop_{0}^{\pi/2}{d\theta \over \sqrt{1-k^2\sin \theta
}},\qquad
E(k) = \intop_0^{\pi/2}\sqrt{1-k^2\sin \theta }\,d\theta,\qquad 0\le k^{2}\le 1
$$

In this way taking sufficiently large $l$ we can choose $\mu$ such that
$\RG>0$ and quite arbitrary~$\z$ ($\uP<0$) to satisfy a zero energy condition
for an excitation of a warp factor.

Taking $m_{\u A}=m\dr{EW}\simeq80\,$GeV, $m\dr{pl}\cong2.4\cdot
10^{18}\,$GeV, $\a_s=\sqrt{\a\dr{em}}=\frac1{\sqrt{137}}$, one gets from
Eq.~(144) 
$$
\mu=-3.581552661\ldots+4.1\cdot10^{-35}\frac{q(\z)}{l^{4/7}} \eqn146
$$
which justifies our approximation for $\e$ and simultaneously gives an
account for a smallness of a \co\ \ct\ $\t R(\t\G(\mu))\simeq0$. Thus it
seems that in this simple toy model we can achieve a large $c\dr{eff}$
without considering large \co\ \ct. 

Let us give some numerical estimation for a time needed to travel a distance
of $200\,$Mps
$$
t\dr{travel}=\frac L{c\dr{eff}\gi{max}}=\frac Lc\cdot e^{-(2l+1)\pi/4}.
$$
One gets for $L=200\,$Mps,
$t\dr{travel}=2\cdot 10^{13}\,\text{s}\cdot e^{-l\pi/2}$
and for $l=100$, $t\dr{travel}\cong12\,$ns.

Let us notice that for $l=100$, $\mu$ from Eq.~(146) is equal to
$$
\mu=-3.581552661\ldots + 5.94\cdot 10^{-44}q(\z). \eqn147
$$
$q(\z)$ cannot be too large (it is a part of \co\ term $q(\z)=|\uP|^{15/7}$).
If $\uP$ is of order $0.1$, $q$ is of order $10^{-16}$; if $\uP$ is of order
$1.1$, $q$~is of order~$0.6$. Thus $\mu$ is very close to~$\mu_0$ for some
reasonable values of~$|\uP|$. It is interesting to ask how to develop this
model to more dimensional case. It means, to the manifold~$M$ which is not a
circle. It is easy to see that in this case we should consider a warp factor
which depends on more coordinates taking under consideration some ansatzes. 
For example we can consider a warp factor in a shape
$$
\exp\Bigl(\sum_{i=1}^m \sqrt{\t A_i}\,f_i(y)\Bigr) \eqn148
$$
where $f_i(y)=x_i$, $i=1,2,\ldots,m=\frac12n_1(n_1+1)$ are parametric
equations of the manifold~$M$ in $m$-dimensional euclidean space and
$$
ds^2_M=\biggl(\sum_{k=1}^m \(\frac{\pa f^k}{\pa y^i}\)
\(\frac{\pa f^k}{\pa y^j}\)\biggr) dy^i \otimes dy^j \eqn149
$$
is a line element of the manifold $M$. In the case of the manifold $S^2$ it
could be
$$
\al
f_1&=\cos\theta\sin\la\cr
f_2&=\cos\theta\cos\la\cr
f_3&=\sin\theta\cr
m&=3
\eal
\eqn150
$$
and a warp factor takes the form
$$
\al
&\exp\(\sqrt{\t A_1}\cos\theta \sin\la+\sqrt{\t A_2}\cos\theta \cos\la
+\sqrt{\t A_3}\sin\theta \), \cr
&\theta\in\langle0,\pi),\ \la\in\langle0,2\pi).
\eal \eqn151
$$

Afterwards we should develop a warp factor in a series of spherical harmonics
on the manifold~$M$ and calculate an energy of a tower of scalar particles
connected with this development. We should of course regularize the series
(divergent) using $\z$-functions techniques. In this case we should use
$\z$-functions connected with the Beltrami-Laplace operator on the
manifold~$M$. Afterwards we should find a zero energy condition for special
types of \ct s $\t A_i$, $i=1,2,\ldots,m$. We can of course calculate the
energy of an excitation of the warp factor using some formulae from classical
field theory. In the case of $M=S^2$ we have to do with an ordinary Laplace
operator on~$S^2$ and with ordinary spherical harmonics with spectrum
given by 
$$
\overline{m}(\ell ,m) = {1\over r}\sqrt{\bigg(1+{\displaystyle{n^2\zeta ^2\over
\overline{M}(\zeta^2+1)}}\bigg)\ell(\ell+1)},\quad\ \ell = 1,2,3.
$$
The Fourier analysis of (151) can be proceeded
on~$S^2$. 

Moreover a toy model with $S^1$ can be extended to the \co\ background in
such a way that 
$$
ds^2_5=e^{\sqrt{\u A}\,\la}\,dt^2 - R^2(t)(dx^2+dy^2+dz^2)-r^2d\la^2. \eqn152
$$

It seems that from practical (calculational) point of view it is easier to
consider a metric of the form
$$
ds^2_5=dt^2-e^{-\sqrt{\u A}\,\la}R^2(t)(dx^2+dy^2+dz^2)- r^2\,d\la^2 \eqn153
$$
where $R(t)$ is a scale factor of the Universe. For further investigations we
should also consider more general metrics, i.e.
$$
\al
ds^2_6&=dt^2-\exp\(-\sqrt{\t A}\bigl(\cos\theta(\sin\la+\cos\la)+\sin\theta\bigr)\)\\
&\times R^2(t)(dx^2+dy^2+dz^2)- r^2\(d\theta^2+\sin^2\theta\, d\la^2\)
\eal \eqn154
$$
or even
$$
\al
ds^2_{4+n_1}&=dt^2-\exp\(-\sum_{i=1}^m \sqrt{\t A_1}f_i(y)\)
\cdot R^2(t)\,\frac{dx^2+dy^2+dz^2}{1-k(x^2+y^2+z^2)}\\
&- r^2\biggl(\biggl(\sum_{k=1}^m \(\frac{\pa f^k}{\pa y^i}\)\(\frac{\pa f^k}
{\pa y^j}\)\biggr)\cdot dy^i\otimes dy^j\biggr)
\eal \eqn155
$$
where
$$
m=\tfrac12\,n_1(n_1+1). \eqn156
$$

Let us notice that our toy model with a travel in a hyperspace (this is
really a travel along dimensions of a vacuum state manifold) can go to some
acausal properties. This is evident if we consider two signals: one sent in
an ordinary way with a speed of light and a second via the fifth dimension to
the same point. The second signal will appear earlier than the first one at
the point. Thus it will affect this point earlier. Effectively it means a
superluminal propagation of signals in a Minkowski space (or in
Friedmann-Robertson-Walker Universe). In this way we can construct a
time-machine using this solution (even if it does not introduce tachyons in a
4-dimensional space-time), i.e.\ a space-time where there are closed
time-like (or null) curves.

Let us notice that given a superluminal signal we can always find a reference
frame in which it is travelling backwards in time. Suppose we send a signal
from~$A$ to~$B$ (at time $t=0$) at an \ef\ speed $u\dr{eff}>1$ in frame~$S_1$
(it means we send a signal through a path described here earlier and an \ef\
speed $u\dr{eff}>1$) with coordinates $(t,x)$ (we consider only one space
dimension). \hbox{In a frame~$S_2$ moving \wrt $S_1$} with velocity
$v>\frac1{u\dr{eff}}$, the signal travels backwards in $t'$ time. This
follows from the Lorentz transformation, i.e.
$$
\ealn{
t'_B&=\frac{t_B(1-vu\dr{eff})}{\sqrt{1-v^2}} &\rf157 \cr
x'_B&=\frac{t_B(1-\frac v{u\dr{eff}})}{\sqrt{1-v^2}}\,. &\rf158
}
$$
We require both $x'_B>0$, $t'_B<0$, that is $\frac1{u\dr{eff}}<v<u\dr{eff}$
which is possible only if $u\dr{eff}>1$.

However this by itself does not mean a violation of a causality. For that we
require that the signal can be returned from~$B$ to a point in the past of
light cone of~$A$. Moreover, if we return the signal from~$B$ to~$C$ with the
same \ef\ speed (it means we send a signal through the fifth dimension with
the same \ef\ speed using the warp factor) in frame~$S_1$, then it arrives
at~$C$ in the future cone of~$B$. The situation is physically equivalent in
the Lorentz boosted frame~$S_2$---the return signal travels forward in
time~$t'$ and arrives at~$C$ in the future cone of~$A$. This is a
frame-independent statement. If a backwards-in-time signal $AC$ is possible
in frame~$S_1$, then a return signal sent with the same speed $u\dr{eff}$
will arrive at a point~$D$ in the past light cone of~$A$ creating a closed
time-like curve $ACDA$. In a Minkowski space (a~space for $\la=0$, or in
general for $\la=\la_0=\text{const.}$), local Lorentz invariance results in
that if a superluminal signal such as $AB$ is possible, then so is one
of~$AC$ for it is just given by an appropriate Lorentz boost. The existence
of a global inertial frames in a Minkowski space guarantees the existence of
the return signal~$CD$.

Thus we get unacceptable closed time loops. Moreover the theory for the
five-dimensional world is still causal, as in five-dimensional General
Relativity. Let us notice that in a \co\ background described by a metric \rf
153 \ the situation will be very similar. We have here also global inertial
frame for $\la=0$, and similar possibility to send a superluminal signal
through fifth dimension. However, in this case we should change a shape of
Lorentz transformation to include a scale factor $R(t)$ which slightly
complicates considerations. Moreover in order to create a warp factor we need
a zero energy condition. If this condition is not satisfied we need an
infinite amount of an energy to create it. The zero energy condition imposes
some restriction on parameters which are involved in our unified theory.
Maybe it is impossible to satisfy these conditions in our world. In this way
it would be an example of Hawkins' chronology protection conjecture: ``{\sl
the laws of physics do not allow the appearance of closed timelike
curves\/}''. However, if a zero energy condition is satisfied in a realistic
Nonsymmetric Kaluza--Klein (Jordan--Thiry) Theory (it means in such a theory
which is consistent with a phenomenology of a contemporary elementary
particle physics) then we meet possibility of \ef\ closed time-like loops
(even the higher dimensions are involved). In this case we should look for
some weaker conjectures (especially in an expanding Universe, not just in a
Minkowski space). Such weaker condition protects us from some different
causal paradoxes as the following paradox: a highly energetic signal can
destroy a laboratory before it was created. In some sense it means some kind
of selfconsistency conditions.

Thus what is a prescription to construct a time-machine in our theory? It is
enough to have a 5-dimensional (or ($n_1+4$)-dimensional) space-time with a
warp factor described here earlier. Afterwards we need two frames $S_1$
and~$S_2$ in relative motion with relative velocity $v>\frac1{u\dr{eff}}$
where $u\dr{eff}$ is an \ef\ superluminal velocity $u\dr{eff}>1$. This could
be arranged e.g.\ as two spaceships with relative motion ($A$~and~$C$).
Sending a signal with \ef\ velocity $u\dr{eff}$ from $A$  to~$C$ we get the
signal in a past of~$C$. Afterwards we send a signal (through the fifth
dimension) from $C$ to~$A$. In this way we arrive in a past of~$A$.

If we want to travel into a future we should reverse a procedure with
$$
-1<v<0.
$$
In both cases we use an \ef\ time-like loop, constructed due to a warp factor
in a five-dimensional space-time. Using this loop several times we can go to
the future or to the past as far as we want. An interesting point which can
be raised consists in an \ef ness of such a travel. It means how much time we
need to get e.g. 1\,s to past or 1\,s to future. It depends on a relative
velocity $v$ and~$u\dr{eff}$. The \ef ness can be easily calculated.
$$
\eta=\frac{u\dr{eff}v}{\sqrt{1-v^2}}\,.\eqn159
$$
If we use our estimation for $c\dr{eff}$ (see Eq.\ \rf123 ), we get
$$
\eta=\frac{v}{\sqrt{1-v^2}}\, e^{(2l+1)\pi/4} \eqn160
$$
for $l$ large. It is easy to see that $\eta\to\infty$ if $v\to1$ and if $l
\to\infty$. 

Let us take $v=\frac1{10}$ and $l=100$. In this case one gets
$$
\eta\approx 10^{67}. \eqn161
$$
What does this number mean? It simply means that in order to go 1\,s into the
past or into the future we loose $10^{-67}$\,s of our life.

However, to be honest, we should use as $u\dr{eff}$ a velocity different from
$c\dr{eff}$. In this case a time needed to travel through the fifth dimension
really matters. Let a speed along the fifth dimension be equal to the
velocity of light (or close to it). In this case one gets
$$
u\dr{eff}=\frac{c\dr{eff}}{1+2(\frac{r_0}L)c\dr{eff}} \eqn162
$$
where $r_0\simeq 10^{-16}$\,cm, $L$ is the distance between $A$ and~$B$,
e.g.\ 1000\,km. In this case $u\dr{eff}$ is smaller,
$$
u\dr{eff}=10^{26} \eqn163
$$
and
$$
\eta\simeq 10^{25}. \eqn164
$$
In this case we can estimate how much time we loose to go 2000 years back or
forward. It is $6.3\cdot10^{-15}$\,s. However, if we want to see dinos we
should go back 100\,mln\,yr and we loose $10^{-10}$\,s. Of course in the
formula for $u\dr{eff}$ we can use smaller~$L$ and a velocity in the place
$c\dr{eff}$ can be smaller too. We can travel more comfortably changing
smoothly a velocity. Any time a time to loose to travel back or forward in
time for our time machine is very small. In some sense it is very \ef\ (if it
is possible to construct).

\def\cor{\gi{corrected}}
Let us notice that the warp factor (or $\u A$) is fixed by the \ct s of the
theory (especially by a \co\ \ct\ $\la_{c0}$). In this way only in some
special cases we get a zero energy condition. However, in the case of the
warp factor existence we have to do with a tower of scalar fields (not only
with a \q). The \q\ field is a zero mode of this tower. Thus in the case of
an existing of a tower of scalar fields we will have to do with a corrected
\co\ \ct, which is really an averaged \co\ term over $S^1$ (in general
over~$M$) 
$$
\la\cor_{c0}=\frac1{2\pi}\,\int_0^{2\pi} \(|\g|-\b e^{2\P(\la)}\)
e^{n\P(\la)}\,d\la \eqn165
$$
where $\P(\la)$ is given by Eq.\ \rf59 . In this way according to our
considerations concerning an energy of excitation
$$
\la\cor_{c0}=E\dr{wf}\,. \eqn166
$$
However, in the formula for $E\dr{wf}$ we dropped an energy for a zero mode
for a tower of scalar fields (i.e., for a \q). Thus
$$
\la\gi{tot}_{c0}=\la_{c0}(\P_0)+\la\cor_{c0}. \eqn167
$$
If the zero energy condition for a warp factor is satisfied, we get
$$
\la\gi{tot}_{c0} =\la_{c0} \eqn168
$$
and we do not need any tunning of parameters in the theory. However, now we
should write down full 5-dimensional equations in our theory ($n_1+4$ in
general) to look for more general solutions mentioned here earlier. This will
be a subject of a future work.

\def\ii#1 {\item{[#1]}}
\section{References}
\setbox0=\hbox{[11]\enspace}
\parindent\wd0

\ii1 {Kalinowski} M. W., 
{\it Nonsymmetric Fields Theory and its Applications\/},
World Scientific, Singapore, New Jersey, London, Hong Kong 1990.

\item{} {Kalinowski} M. W., 
{\it Can we get confinement from extra dimensions}\/,
in: Physics of Elementary Interactions (ed. Z.~Ajduk, S.~Pokorski,
A.~K.~Wr\'oblewski), World Scientific, Singapore, New
Jersey, London, Hong Kong 1991.

\item{} {Kalinowski} M. W., 
{\it Nonsymmetric Kaluza--Klein (Jordan--Thiry) Theory in a general
nonabelian case}\/,
Int. Journal of Theor. Phys. {\bf30}, p.~281 (1991).

\item{} {Kalinowski} M. W., 
{\it Nonsymmetric Kaluza--Klein (Jordan--Thiry) Theory in the electromagnetic
case}\/,
Int. Journal of Theor. Phys. {\bf31}. p.~611 (1992).

\ii2 {K{\"a}lbermann} G., 
{\it Communication through an extra dimension}\/,
Int.~J. of Modern Phys.~{\bf A15}, p.~3197 (2000).

\item{} {Visser} M.,
{\it An exotic class of Kaluza-Klein model}\/,
Phys. Lett. {\bf B159}, p.~22 (1985).

\item{} {Li} Li-Xin, {Gott} III J. R.,
{\it Inflation in Kaluza-Klein theory: Relation between the fine-structure
constant and the cosmological constant}\/,
Phys. Rev. {\bf D58}, p.~103513-1 (1998).

\item{} {Squires} E. J.,
{\it Dimensional reduction caused by a cosmological constant}\/,
Phys. Lett. {\bf B167}, p.~286 (1986).

\item{} {Caldwell} R., {Langlois} D.,
{\it Shortcuts in the fifth dimension}\/,
Phys. Lett. {\bf B511}, p.~129 (2001).

\ii3 {Dick} R.,
{\it Brane worlds}\/,
Class. Quantum Grav. {\bf18}, p.~R1 (2001).

\item{} {Brax} Ph., {Falkowski} A., {Lalak} Z.,
{\it Non-BPS branes of supersymmetric brane worlds}\/,
Phys. Lett. {\bf B521}, p.~105 (2001).

\ii4 {Randall} L., {Sundrum} R.,
{\it An alternative to compactification}\/,
hep-th//9906064.

\ii5 {Arkami-Hamed} N., {Dimopoulos} S., {Dvali} G.,
{\it The hierarchy problem and new dimensions at a milimeter}\/,
Phys. Lett. {\bf B429}, p.~263 (1998).

\end